\documentclass[final,5p,times,twocolumn]{elsarticle}
\usepackage{amsmath}
\usepackage{geometry}
\usepackage{float}
\usepackage{color}
\usepackage{soul}
\usepackage[utf8]{inputenc}
\usepackage{graphicx}

\def\etal{\mbox{\it et al.\ }}

\usepackage{multirow}

\journal{}

\begin{document}

\begin{frontmatter}

\title{Exploration of the High Entropy Alloy Space as a Constraint Satisfaction Problem}
\author[msen]{A. Abu-Odeh}
\author[meen]{E. Galvan}
\author[meen]{T. Kirk}
\author[thsw,kth]{H. Mao}
\author[thsw]{Q. Chen}
\author[tcus]{P. Mason}
\author[meen]{R. Malak}
\author[msen,meen]{R. Arr\'{o}yave\corref{cor1}}
\ead{rarroyave@tamu.edu}

\address[msen]{Department of Materials Science and Engineering, Texas A\&M University, College Station, TX 77843-3123, USA}
\address[meen]{Department of Mechanical Engineering, Texas A\&M University, College Station, TX 77843-3123, USA}
\address[thsw]{Thermo-Calc Software AB, R{\aa}sundav\"{a}gen 18A, SE169 67 Stockholm, Sweden}
\address[kth]{Materials Science and Engineering, KTH Royal Institute of Technology, SE 100 44  Stockholm, Sweden}
\address[tcus]{Thermo-Calc Software Inc, McMurray, PA, 15317, USA}

\cortext[cor1]{Corresponding author}

\begin{abstract}
High Entropy Alloys (HEAs), Multi-principal Component Alloys (MCA), or Compositionally Complex Alloys (CCAs) are alloys that contain multiple principal alloying elements. While many HEAs have been shown to have unique properties, their discovery has been largely done through costly and time-consuming trial-and-error approaches, with only an infinitesimally small fraction of the entire possible composition space having been explored. In this work, the exploration of the HEA composition space is framed as a Continuous Constraint Satisfaction Problem (CCSP) and solved using a novel Constraint Satisfaction Algorithm (CSA) for the rapid and robust exploration of alloy thermodynamic spaces. The algorithm is used to discover regions in the HEA Composition-Temperature space that satisfy desired phase constitution requirements. The algorithm is demonstrated against a new (TCHEA1) CALPHAD HEA thermodynamic database. The database is first validated by comparing phase stability predictions against experiments and then the CSA is deployed and tested against design tasks consisting of identifying not only single phase solid solution regions in ternary, quaternary and quinary composition spaces but also the identification of regions that are likely to yield precipitation-strengthened HEAs. 
\end{abstract}


\end{frontmatter}


\section{Introduction}
\label{sec:intro}
\subsection{Motivation}
High Entropy Alloys (HEAs), Multi-principal Component Alloys (MCA), or Compositionally Complex Alloys (CCAs) are alloys that contain multiple principal alloying elements that differ from traditional engineering alloys in that they are not based on a major constituent but instead are located closer to the center of the composition space~\cite{miracle2017critical,gao2016high}. The “High Entropy” qualifier was overwhelmingly used in the past to emphasize––what was once thought as––the major defining characteristic of such alloy systems, namely the existence of an extended single phase solid solution stabilized through configurational entropy. However, “entropy stabilization” has gradually lost relevance due to the fact that (i) it is not really clear that configurational entropy plays a determinant stabilizing role~\cite{miracle2017critical}; (ii) most of the recent emphasis has shifted towards identification of multi-phase HEA/CCAs with unique multi-phase microstructural features~\cite{miracle2015critical,he2016precipitation}.

While it is certainly true that out of the hundreds of HEA/CCAs so far investigated––out of a truly vast composition-microstructure space~\cite{miracle2017critical}––, many exhibit properties that are comparable and often worse than conventional alloys sitting at corners of the composition space~\cite{miracle2017critical}, it is also true that some alloys have shown remarkable properties. In general, HEA/CCAs containing passivating elements such as Cr~\cite{shang2017high,shi2017corrosion} exhibit equivalent or superior corrosion-resistant properties compared to conventional alloys~\cite{chen2005microstructure}. Furthermore, some HEA/CCAs exhibit exceptional properties, such as combined strength-ductility performance~\cite{senkov2014effect,li2016metastable}, improved fatigue resistance~\cite{hemphill2012fatigue,tang2015fatigue}, high fracture toughness~\cite{seifi2015fracture}, and high thermal stability. Response is even more remarkable when exploiting strengthening effects due to controlled precipitation of coherent secondary phases~\cite{tsao2017superior}, and grain refinement. 

\subsection{On the Efficient Exploration of HEA Space}
Within a decade, the exploration of the composition-microstructure-property space in HEA/CCA systems has grown dramatically, with multiple groups and concerted efforts around the world being dedicated to the full exploitation of this new concept in alloy design. Over time, several alloys with outstanding properties have been discovered but in truth only a few hundred individual alloys have been investigated~\cite{miracle2017critical}. When considering the vastness of the HEA space, with literally uncountable possible combinations of four, five, six and more elements within arbitrary ranges of composition, it is clear that it is necessary to develop accelerated strategies for its efficient exploration.

The original premise of research activities on HEAs was that configurational entropy is maximal in solid solutions with five or more elements at nearly equiatomic concentrations~\cite{miracle2017critical}, stabilizing random solid solution states relative to other more ordered competing phases and avoiding miscibility gaps. Real systems, however, are never likely to form truly random solid solutions as there are always tendencies to either order or phase-separate depending on the nature of the chemical interactions between constituents (exothermic vs endothermic). Thus, Short Range Order (SRO) contributions need to be taken into account. van de Walle \etal show, for example, how SRO in fact stabilizes random solutions against ordered states~\cite{van2017software} since the reduction in configurational entropy in the solid solution due to SRO is more than compensated by the exothermic enthalpic interactions enhanced by SRO. 

Moreover, the focus on (random-like) configurational entropy sometimes ignores other (excess) contributions to the entropy of an alloy (electronic, vibrational, magnetic) that contribute to the total entropy and that may not necessarily cancel out when solid solutions are more ordered (less 'entropy-stabilized'). More problematic is the fact that relating maximal entropy to maximal stability ignores the fact that under prevalent isothermal-isobaric conditions the Gibbs energy ($G=H-TS$) is the potential that defines---through its minimization---the equilibrium state of a chemical system.

\subsection{Searching for Predictors for Solid Solutions}
Regardless of whether the focus on configurational entropy is warranted from the stand point of phase stability, numerous efforts over the years have focused on the identification of features related to the likely existence of a solid solution at arbitrary (mostly equiatomic) alloy compositions. Early on, Zhang \etal~\cite{zhang2008solid} proposed the use of entropy of mixing, $\Delta S_{mix}$, atomic size mismatch, $\delta$ as well as enthalpy of mixing $\Delta H_{mix}$ to predict existence of HEAs. Essentially, large $\Delta S_{mix}$, small (in magnitude) $\Delta H_{mix}$ and $\delta$ would be conducive to the formation of random solid solutions (RSS). Later on, Guo \etal~\cite{guo2011effect,sheng2011phase} proposed the use of the valence electron concentration ($VEC$) as a discriminator between FCC-forming and BCC-forming HEAs, in the spirit of early work on crystal structure classification efforts~\cite{paxton1997bandstructure}. Guo \etal~\cite{sheng2011phase} also proposed to use the mismatch in electronegativity $\Delta \chi$ as an indicator for the likelihood of solid solution.

Yang \etal later proposed the use of the so-called $\Omega$ parameter:
\begin{equation*}
\Omega = \frac{T_m\Delta S_{mix}}{| \Delta H_{mix}|}>1
\end{equation*} 
\noindent based on an analysis of the competition between entropic and enthalpic contributions to the free energy of a phase. The analysis suggested that when $\Omega  > 1$ there was a greater chance for stable random solid solutions relative to either phase separation or intermetallic formation. Poletti \etal~\cite{poletti2014electronic} expanded the thermodynamic analysis by considering the chemical stability of a random solution (from an analysis of the sign of the $|\frac{\partial^2 G}{\partial x_i \partial x_j}|$ matrix) introducing the $\mu$ parameter:
\begin{equation*}
\mu = \frac{T_{m}}{T_{SC}}>>1
\end{equation*}
\noindent where $T_m$ is the weighed average melting temperature of the alloy and $T_{SC}$ is the critical temperature of the miscibility gap, with larger values of $\mu$ favoring solid solutions. Later, Senkov \etal~\cite{senkov2016new} introduced a criterion, $\kappa_1^{cr}\left(T\right)$, based on the (enthalpic) competition between intermetallic compounds and solid solutions:
\begin{equation*}
\kappa_1^{cr}\left(T\right) > \frac{\Delta H_{IM}}{\Delta H_{mix}}
\end{equation*}

Toda-Caraballo and Rivera D\'{i}az del Castillo~\cite{toda2016criterion} reviewed prior attempts at identifying features capable of predicting the phase constitution state of a given composition. They concluded that neither feature set was capable of properly classifying the likely phase constitution of an arbitrary (equiatomic) alloy and they proposed instead the use of two additional parameters, mismatch in interatomic distance, $s_m$ and bulk modulus, $K_m$. Testing the performance of $s_m$, $K_m$ as well as other parameters introduced earlier it was found that while the parameters were indicative, there was considerable overlap in the regions in the $s_m$-$K_m$ space corresponding to FCC/HCP/BCC solid solutions, solid solutions + intermetallics, etc. This is clearly insufficient when attempting to design HEAs with specific phase constitution characteristics.

While all the parameters proposed in the past have some physical/thermodynamic grounding, no combination of two parameters has been shown to provide sufficiently robust predictions for the likely phase constitution of a given alloy chemistry~\cite{tancret2017designing}. Using simultaneously more than a few parameters at a time in order to achieve better predictive ability has been explored with varying degrees of success. Dom\'{i}nguez \etal~\cite{dominguez2015prediction}, for example, proposed an approach in which statistical learning techniques (specifically Principal Component Analysis, PCA) were used against thermodynamic and electronic properties of HEAs to distinguish between BCC and FCC structures. 

Following the same ideas by Dom\'{i}nguez \etal, Tancret \etal~\cite{tancret2017designing} have put forward a novel approach in which they combine alloy indicators~\cite{toda2016criterion} with computational thermodynamics information derived from CALPHAD databases to develop robust predictors for single phase solid solutions. Their approach establishes a probabilistic measure (based on a Gaussian Process predictor) of a given alloy with specific composition in terms of its likelihood to form either a single phase solid solution, or a multi-phase alloy with a majority phase being a solid solution phase. While this framework for the classification of multi-component alloys has been shown to be successful, it is not clear whether such an approach can truly be used to design HEAs.

\subsection{Thermodynamics-based Exploration of the HEA Space}
As discussed above, the search for unique identifiers for single-phase solid solutions has had a rather limited success. This could be partially explained because of the lack of sufficient experimental information to develop robust predictors as the existing data ( a few hundred alloys) is infinitesimally small relative to the vast, multi-dimensional HEA space. There is, however, a more fundamental limitation to such approaches since they treat a problem of alloy stability as an intrinsic property of a material. Phase stability results from the competition among multiple phases to form a state with \emph{minimal total Gibbs energy} at a specific temperature and pressure,  subject to mass conservation and other appropriate constraints. Since competition among phases is the result of Gibbs energy differences that are usually less than a \emph{ few tens of meV/atom (or kJ/mol)}, it is unlikely that coarse measures of enthalpic or entropic contributions to the Gibbs energy of one phase would yield robust estimates of the phase stability of a complex multi-component, multi-phase system. Indeed, perhaps the most promising approaches to date when it comes to exploration of the HEA alloy space have been based on approaches that predict phase stability as the result of competitions among multiple phases that may form an equilibrium thermodynamic state.

Phase stability analysis using DFT-based energetics~\cite{troparevsky2015criteria,troparevsky2015beyond} have increased in popularity due to the increased availability of \emph{ab initio} databases. Troparevsky~\cite{troparevsky2015criteria}, for example, used enthalpies of formation for the binary subsystems comprising candidate HEA alloys available in \emph{ab initio} databases to estimate whether a given multi-component was likely to exhibit solid solution behavior by comparing those enthalpies with the expected (maximum) configurational entropy, although they did not consider finite-temperature contributions to the free energies of the competing phases. 

Recently, Wang \etal~\cite{wang2018computation} carried out fully quasi-harmonic calculations on the MoNbTaVW system, accounting for phonon- and electron-thermally excited contributions to the free energy of several unary, binary and higher-order ordered compounds. Using then the Simplex algorithm at different temperatures they were able to identify the decomposition reactions for the quinary and the different ternary and quaternary subsystems, predicting decomposition temperatures consistent with available experiments. The search was over a single quinary system and they concluded that most subsystems were likely to decompose into ordered phases at temperatures below those observable experimentally.

Lederer \etal~\cite{lederer2017search} recently carried a \emph{tour the force} investigation of the phase stability in 130 quinary, 1100 quaternary and over 4,000 ternary systems by combining an \emph{ab initio} database of binary and ternary compounds with cluster expansion calculations on BCC and FCC lattices as well as a mean-field statistical mechanical model to predict the temperatures of stability for solid solutions (resulting from order/disorder or phase separation/disorder transitions), finding very good agreement with available experimental as well as computational data (derived from CALPHAD calculations). While the method is powerful, it was only applied to equimolar compositions and was limited to predicting the temperature at which solid solutions were expected to be stable.\\

Explicit, finite-temperature prediction of phase equilibrium through computational thermodynamics based on the CALPHAD method is perhaps the more robust approach to the exploration of the HEA phase stability space as these methods allow the prediction of phase equilibria at arbitrary compositions and temperatures~\cite{zhang2014understanding,senkov2015accelerated,senkov2015acceleratedb}. Furthermore, CALPHAD models are parameterized in a hierarchical manner, making it relatively simple to construct thermodynamically-consistent descriptions of higher-order systems from low-order ones. Recently Senkov \etal~\cite{senkov2015accelerated,senkov2015acceleratedb} proposed a high throughput (HT) approach to the exploration of the HEA space in which they carried out CALPHAD-based predictions of phase stability in a large number ($>$130,000) of \emph{equimolar} alloy systems. Through their analysis, they were able to identify promising alloy systems based on specific filtering criteria. A major finding from their work was the fact that their analysis suggested that solid solution alloys become increasingly rare as the number of components increase, which can be interpreted in terms of an increase likelihood of finding pairs of constituents with large exothermic or endothermic interactions that in turn lead to either ordering or phase separation, as independently suggested by Troparevsky~\cite{troparevsky2015criteria} \etal.

\section{Towards the Design of HEAs}
\subsection{The Need for the Goal-Oriented Search of the HEA Thermodynamic Space}
\begin{figure}
\begin{centering}
\includegraphics[width=0.9\columnwidth]{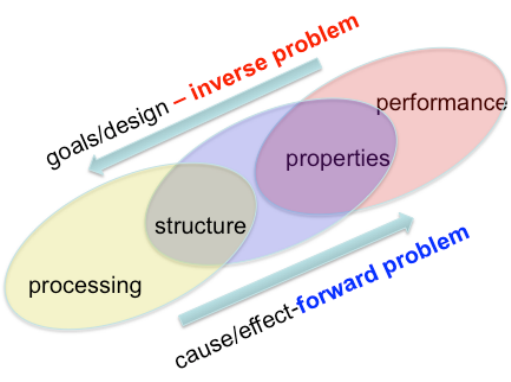}
\par\end{centering}
\caption{Forward vs. inverse modes of materials development~\cite{olson1997computational}.}
\label{fig:goals}
\end{figure}

The central paradigm in materials science is the existence of process-structure-property-performance (PSPP) relationships––Fig.~\ref{fig:goals}. ICME frameworks~\cite{national2008integrated} emphasize the need to establish modeling and simulation tools that act as linkages along this PSPP causal chain. These linkages in turn facilitate a \emph{forward} mode of exploration of the materials space. Design, however, is an inverse, goal-oriented problem and when applied to materials, this can only be achieved by inverting the PSPP paradigm~\cite{olson1997computational}.
Most recent approaches towards the computationally-enabled discovery of HEAs have relied on either the use of ad-hoc alloy design rules, DFT-based predictions of phase stability, or high-throughput CALPHAD calculations. These methods are limited because (i) ad-hoc rules have limited predictive power; DFT-based methods tend to ignore finite-temperature contributions to the free energies of phases that play a fundamental role in phase stability; (ii) high-throughput CALPHAD calculations are not targeted and are incapable of providing compact representations of arbitrary phase stability conditions. All methods above have two further limitations in common as they tend to \emph{focus on determining whether a given \textbf{equiatomic} composition} is likely to yield a single-phase solid solution and are \emph{not targeted or \textbf{goal oriented}}. At the time of this writing, however, we have become aware of a recent paper by Menou \etal~\cite{menou2018evolutionary} where they have used Genetic Algorithm-based optimization subject to constraints to identify regions in the HEA space with optimal mechanical properties. A notable aspect of this work is that they have combined prediction of phase stability with model estimates for strengthening (and other attributes) to carry out a global search of the HEA space. More impressive is the fact that their predictions were corroborated by experiments.

The focus on equiatomic single-phase solid solutions considerably limits the potential alloy design space as it has become evident that, as it has always been the case for traditional structural alloys, real opportunities for improved performance in the HEA space are likely to arise from compositionally––i.e. the nonlinear alloy concept~\cite{miracle2017high}––and microstructurally complex~\cite{miracle2017critical}––e.g. multi-phase HEAs–– systems. Clearly, the search for compositionally/microstructurally complex HEAs implicitly acknowledges a goal-oriented exploration of the HEA space and such a search cannot be done randomly, even through conventional HT approaches. At the phase stability level, this quest to design compositionally/microstructurally complex HEAs is reduced to the problem of identifying the set of thermodynamic conditions that result in \emph{a priori} specified phase stability/constitution requirements. Within the context of \emph{forward}/\emph{inverse} problems along the PSPP paradigm, one could consider standard phase stability calculations/predictions as the \emph{forward} phase stability problem. On the other hand, the targeted exploration of a potentially high-dimensional thermodynamic space can be construed as an \emph{inverse} phase stability problem~\cite{arroyave2016inverse,galvan2017constraint}.

\subsection{The Inverse Phase Stability Problem as a Continuous Constraint Satisfaction Problem}
\begin{figure}
\begin{centering}
\includegraphics[width=0.9\columnwidth]{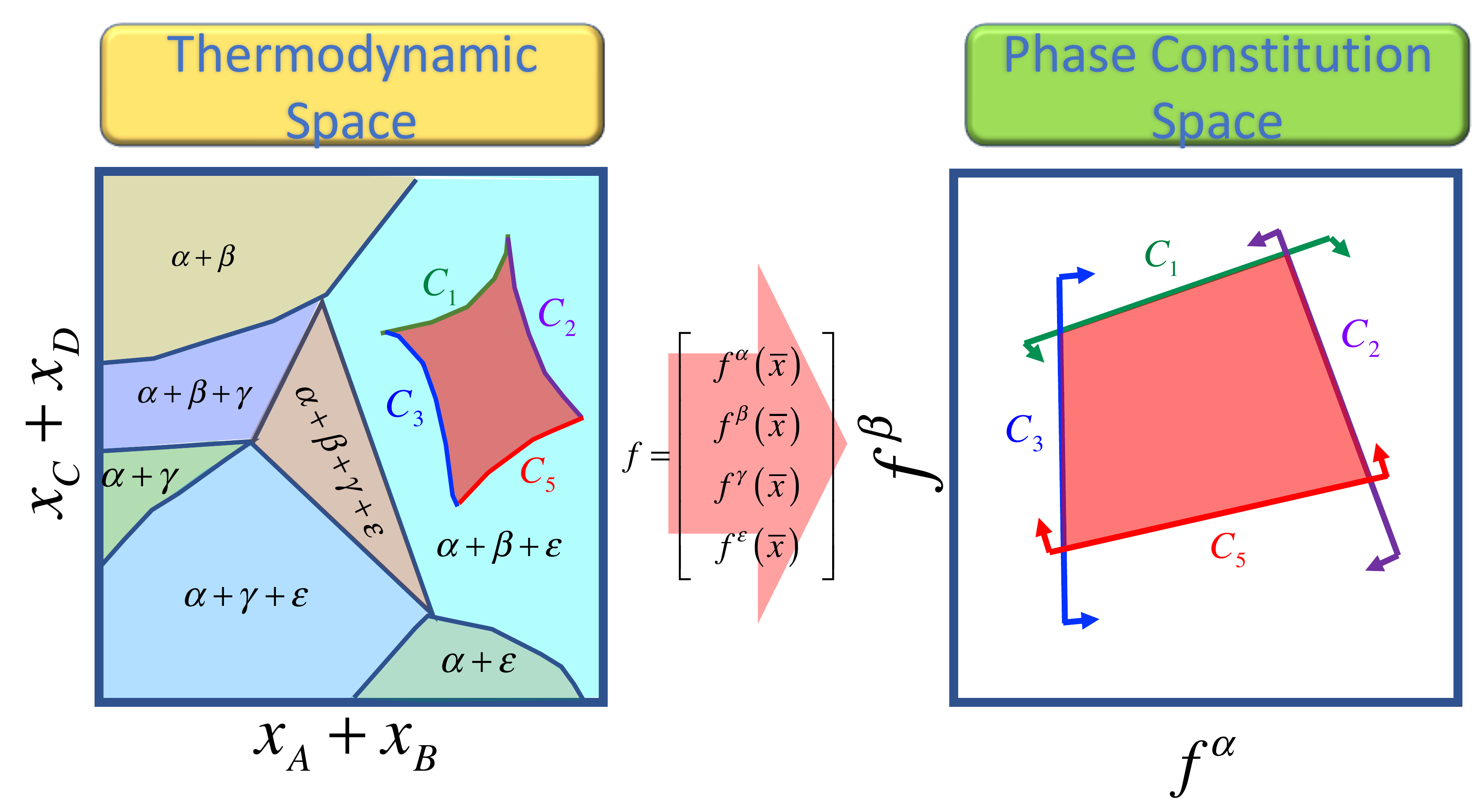}
\par\end{centering}
\caption{Graphical representation of the mapping of the Inverse Phase Stability Problem to a Constraint Satisfaction Problem (CSP). $f$ represents the minimization of the total Gibbs energy of a system. The set of constraints $C=\left(C_1, C_2, C_3, C_4\right)$ is defined by the materials designer in the phase constitution space. Due to the non-linear mapping between the thermodynamic conditions and phase constitutions, the solution in the thermodynamic conditions space may be non-convex~\cite{galvan2014constraint,arroyave2016inverse,galvan2017constraint}.}
\label{fig:csp}
\end{figure}

The \emph{inverse phase stability problem} (IPSP),  consists of identifying the thermodynamic conditions that satisfy desirable phase constitutions that can be expressed in terms of constraints. For example, one could set the goal to identify the entire set of composition-temperature coordinates, $\left(C-T\right)$ that results in a \textbf{single phase} in a given \emph{multi-component} alloy system. In ~\cite{galvan2014constraint,arroyave2016inverse,galvan2017constraint} we have identified this problem to a so-called  Constraint Satisfaction Problem (CSP)~\cite{tsang1995foundations}, which is commonly encountered in a wide range of fields, ranging from Operations Research to Robotics. More specifically, the solution to the IPSP consists of identifying not only \emph{any} but rather \emph{all} conditions that satisfy the set of constraints. This variant of CSPs are known as \emph{Continuous} Constraint Satisfaction Problems (CCSP)~\cite{cruz2005constraint}. The solution to the IPSP thus consists of the set of all points in the thermodynamics condition space that can be mapped to the (constrained) phase constitution space as shown in Fig.~\ref{fig:csp}.

Most techniques developed to solve CCSPs are based on interval arithmetic, branch and bound, or the root inclusion test. However, often these techniques require an analytical expression to determine if a subregion of the search space contains a feasible solution~\cite{hu2014searching}. Since the phase-stability space is non-analytical---phase boundaries represent abrupt transitions between the presence and absence of specific phases---these methods cannot be used for the solution to the IPSP. In recent work, we have developed a mathematically rigorous framework for tackling IPSP using a novel constraint satisfaction algorithm (CSA) based on machine learning techniques~\cite{galvan2014constraint,arroyave2016inverse,galvan2017constraint}.

We note that the idea of framing alloy design as a constraint satisfaction problem has recently been explored independently by Larsen \etal~\cite{larsen2017alloy}. The problem that they wanted to solve was the so-called \emph{inverse lattice problem} in which given a broad class of potentials the challenge is to identify the ground states for \emph{all possible} values of the effective cluster interaction energies used to parameterize the Hamiltonian of the model. A constraint satisfaction model is used to identify constructible configurations. The approach thus identified sets in the model space––i.e. values of configuration interaction parameters––consistent with the constraint that a stable configuration could be constructed from the set. Our work and Larsen \etal constitute the first examples in which inverse problems in materials science have been shown to be framable as constraint satisfaction problems. We note, however, that CSPs are a subclass of a much larger family of problems involving constraints and point, for example, to the work of Tancret and others~\cite{tancret2012computational,menou2016multi} whereby they have used constrained global optimization schemes in alloy design problems.

\begin{figure}
\begin{centering}
\includegraphics[width=0.9\columnwidth]{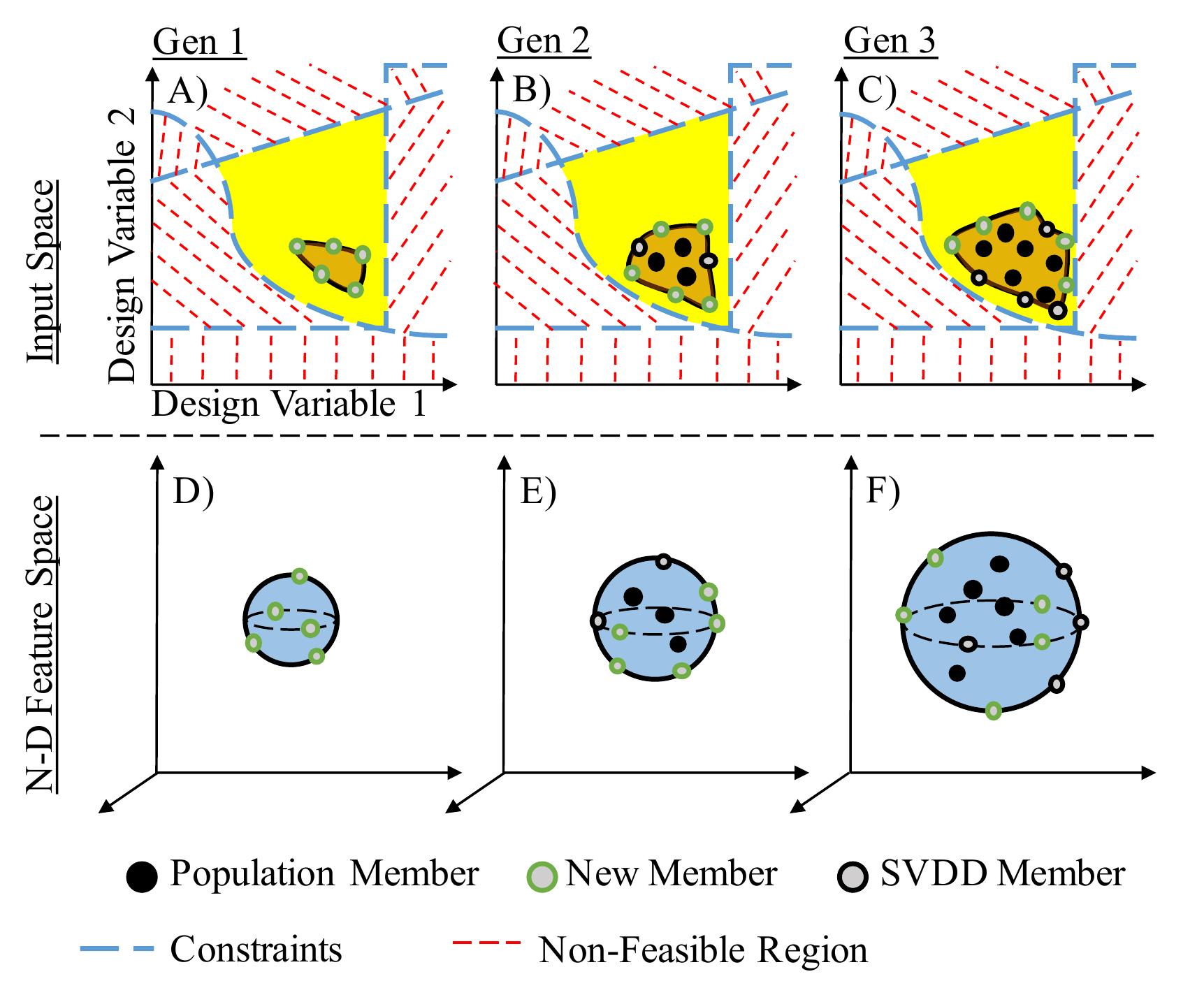}
\par\end{centering}
\caption{The Constraint Satisfaction Algorithm (CSA) at a glance: the first generation (\textbf{A}) of solutions (that satisfy the entire set of constraints) is identified through random exploration of the design (input) space. A support vector domain description (SVDD) is built (\textbf{D}) to represent the CSA solution in the N-D feature space. A Genetic Algorithm (GA)-based optimization scheme is used to expand the radius of the SVDD hyper-sphere (\textbf{E}), which corresponds to an expanded region in the design space (\textbf{B}). The algorithm continues expanding the set of solutions (\textbf{C,F}) to the CCSP until termination criteria is reached~\cite{galvan2014constraint,arroyave2016inverse,galvan2017constraint}.}
\label{fig:csa}
\end{figure}

The CSA–––see Fig.~\ref{fig:csa}–––begins first by randomly exploring the phase stability space. Points in this space that satisfy \emph{all} \emph{a priori} established constraints form a finite, discrete set, which is then generalized into an infinite set through a Support Vector Domain Description (SVDD)~\cite{tax1999support}, which is a machine learning technique. In the context of the CSA, the SVDD acts as a one-class classifier that tags a region in the multi-dimensional space as satisfying the phase stability constraints. Under SVDD, one finds the hypersphere of minimum radius that contains a set of $N$ data points. However, the hypersphere is generally a poor representation of the domain and a kernel function is used to \emph{nonlinearly map} the space data into a higher-dimensional feature space where a hypersphere is a good model~\cite{galvan2014constraint}. 

The data that lie on the hypersphere boundary are called support vectors and they are used to construct the domain description. The SVDD can be constructed in an incremental/decremental manner~\cite{poggio2001incremental}, allowing a relatively inexpensive update of the description as new members are added or removed from the SVDD~\cite{roach2011improved}. The CSA then expands the boundary of the SVDD through either using Genetic Algorithms (GAs) or by using search approaches based on the bi-section method~\cite{galvan2017constraint}.

 We would like to point out that we have shown that the CSA approach is much more efficient than grid-search approaches~\cite{senkov2015accelerated,senkov2015acceleratedb} as the CSA focuses its sampling in regions that are likely to be \emph{most informative} for constructing the constraint boundary model. Grid searching, on the other hand, wastes many samples in regions that provide minimal information about the satisfaction of the constraints.
Upon completion, the SVDD can be re-mapped to the materials design space to compactly and efficiently represent all the regions that satisfy the imposed constraints. This information can then be used to limit the computational/experimental space that needs to be explored for further alloy development. The main advantage of the CSA, however, is the fact that it enables the \textbf{targeted exploration of alloy spaces in order to identify regions of arbitrarily complex phase constitution characteristics}.
 
In practice, the CSA is implemented on top of a Gibbs energy minimization engine that can be used to evaluate if a given set of thermodynamic conditions satisfies or not the constraints. In this work,  the constraints (more about this below) were evaluated through the Thermo-Calc MATLAB API and the Gibbs energy minimization was carried out using Thermo-Calc's High Entropy Alloy database TCHEA1~\cite{mao2017tchea1,chen2017database}. 

\section{Validation of TCHEA1}
Regardless of the rigor of the already developed CSA approach for the exploration of alloys spaces, the effectiveness of the framework ultimately depends on the quality of the phase stability predictions. Recently, the ability of CALPHAD-based frameworks to accurately predict the equilibrium state in HEAs has been put to the test by several groups. Senkov \etal~\cite{senkov2015accelerated,senkov2015acceleratedb} used high-throughput CALPHAD calculations to systematically explore the HEA space. They assembled a collection of thermodynamic databases and assessed their validity based on the fraction of the required binary descriptions included in the database, concluding that a database would have to have descriptions of \emph{at least} all the binaries in order to be deemed as potentially reliable. Gao \etal~\cite{gao2016senary} recently used CALPHAD calculations to examine the phase stability in HfNbTaTiVZr and found good qualitative agreement with experiments. Saal \etal~\cite{saal2018equilibrium} found god agreement between CALPHAD predictions of phase stability and experiments, \emph{provided experiments accounted for the long annealing times necessary to approach equilibrium.} Questions remain, however, as to the validity of CALPHAD databases when it comes to predictions of phase stability in central regions of the composition coordinate system~\cite{miracle2017critical}.

\begin{figure}
\begin{centering}
\includegraphics[width=0.9\columnwidth]{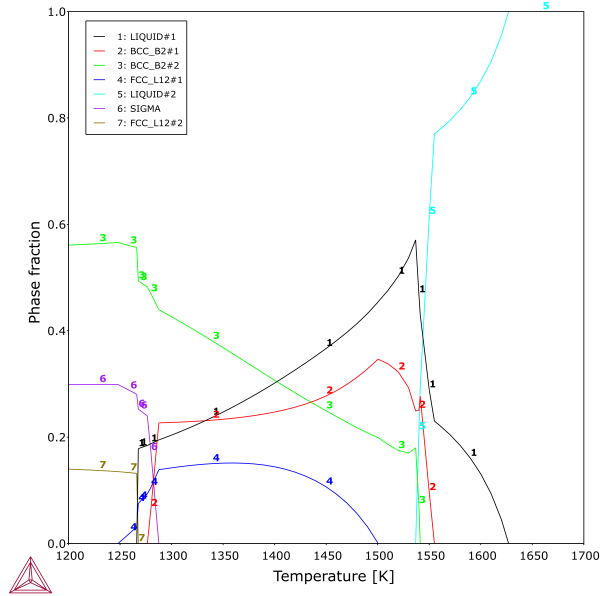}
\par\end{centering}
\caption{Calculated mole fraction of equilibrium phases at various temperatures in the AlCoCrCuFeNi senary alloy with equi-atomic ratio.}
\label{fig:alcocrcufeni}
\end{figure}

\begin{figure}
\begin{centering}
\includegraphics[width=0.9\columnwidth]{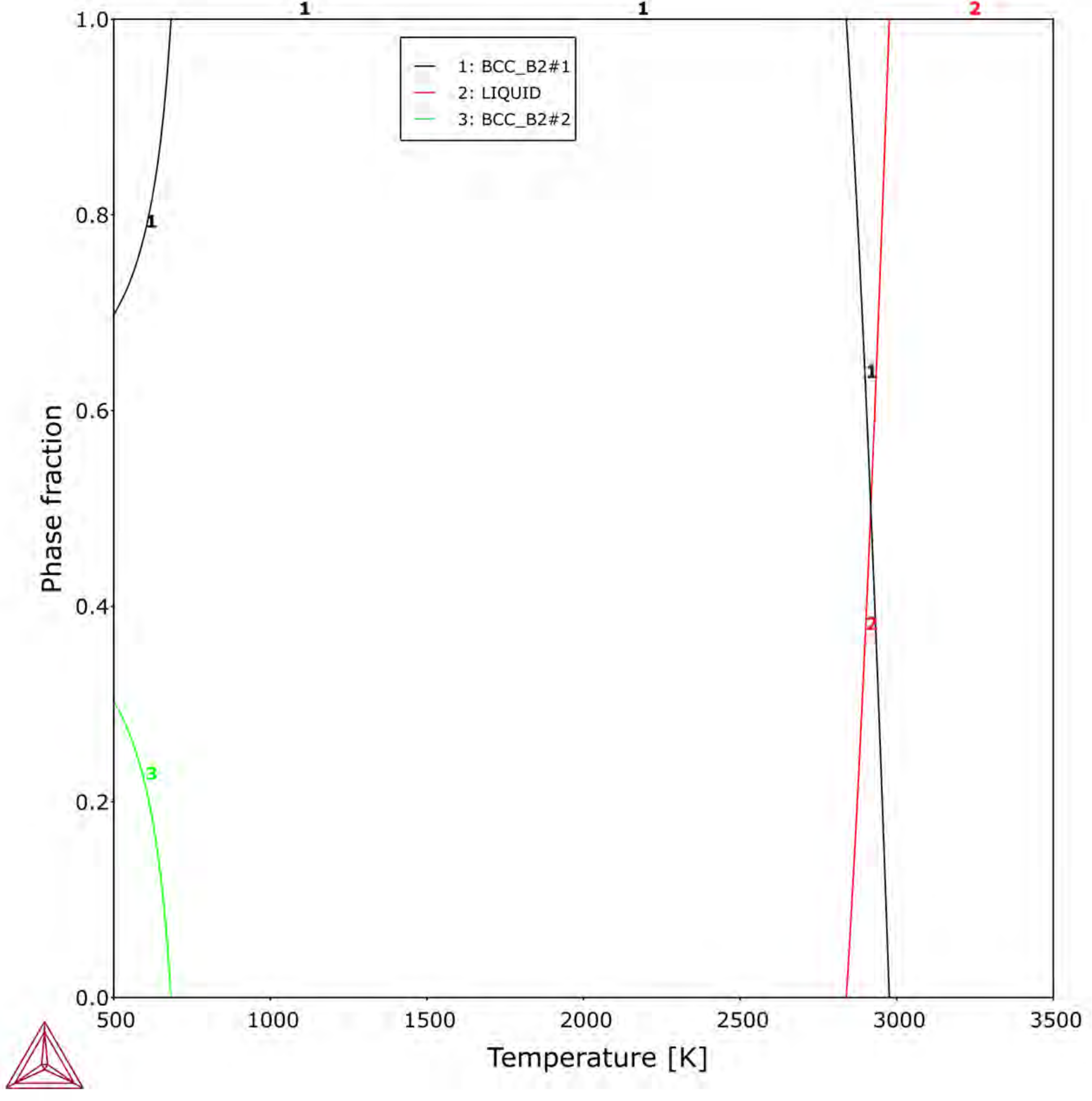}
\par\end{centering}
\caption{Calculated mole fraction of equilibrium phases at various temperatures in the MoNbTaVW quinary alloy with equiatomic compositions.}
\label{fig:MoNbTaVW}
\end{figure}

Very recently, a subset of the present authors~\cite{mao2017tchea1,chen2017database} have carried out in-depth evaluation of the thermodynamic database used in this work, TCHEA1 in the context of HEA phase stability. TCHEA1 describes a 15-element system and includes all the binary subsystems ($\sim$100) and hundreds of ternaries ($>$200). In ~\cite{mao2017tchea1}, Mao \etal considered several synthesized quaternary, quinary, senary and higher order systems. Fig. ~\ref{fig:alcocrcufeni} shows for example, the calculated mole fraction of the equilibrium phases at different temperatures in the AlCoCrCuFeNi senary system (at equiatomic composition), investigated experimentally previously~\cite{li2008effects,tung2007elemental}. Calculations confirmed the duplex FCC+BCC in the as-cast samples, although they suggest that the sigma phase may precipitate in the very late stage of solidification. Sluggish kinetics were rationalized as the reason for the absence of this phase. Better agreement with observations~\cite{senkov2010refractory} was found in the case of the MoNbTaVW quinary system, shown in Fig.~\ref{fig:MoNbTaVW}, which is a prototypical refractory HEA and has been shown both experimentally~\cite{senkov2010refractory} and computationally~\cite{mao2017tchea1} to exhibit a wide solid solubility range.

While similar favorable comparisons were found with several other experimentally synthesized and characterized HEAs, Mao \etal found some instances that exhibited the limitations in the existing database––as of this writing a new version of TCHEA (TCHEA2) is already available with a larger number of elements and updated descriptions on some higher order systems. Furthermore, comparisons with experiments are complicated as it is often the case that reports on HEA are based on as-cast configurations that are usually far from equilibrium, making the comparison with equilibrium states problematic. For example, while the prototypical so-called "Cantor" alloy, CoCrFeMnNi had historically been reported as a stable solid solution over a wide temperature range, it has recently been shown to be unstable against the precipitation of $\sigma$ phase at 700$^0$C after prolonged annealing, in perfect agreement with CALPHAD predictions~\cite{chen2017database}.

Clearly, a much comprehensive comparison needs to be undertaken in order to establish a baseline of confidence in the phase stability predictions based on the Gibbs energy parameterizations of TCHEA1 (or any other database). Recently, Tancret \etal~\cite{tancret2017designing} carried out a systematic evaluation of CALPHAD databases assembled by Thermo-Calc. The databases examined––contrary to TCHEA1––were not explicitly designed to explore the HEA space but in general contained a large number of low order systems. In their work, they examined $>250$ experimentally synthesized alloys and compared them with phase stability predictions using different databases. Overall, Tancret \etal~\cite{tancret2017designing}  reported acceptable but not ideal predictability in their phase stability calculations, although it is to be noted that the databases evaluated in that work were not designed to explore the concentrated alloy space as was TCHEA1.

\begin{figure}
\begin{centering}
\includegraphics[width=0.9\columnwidth]{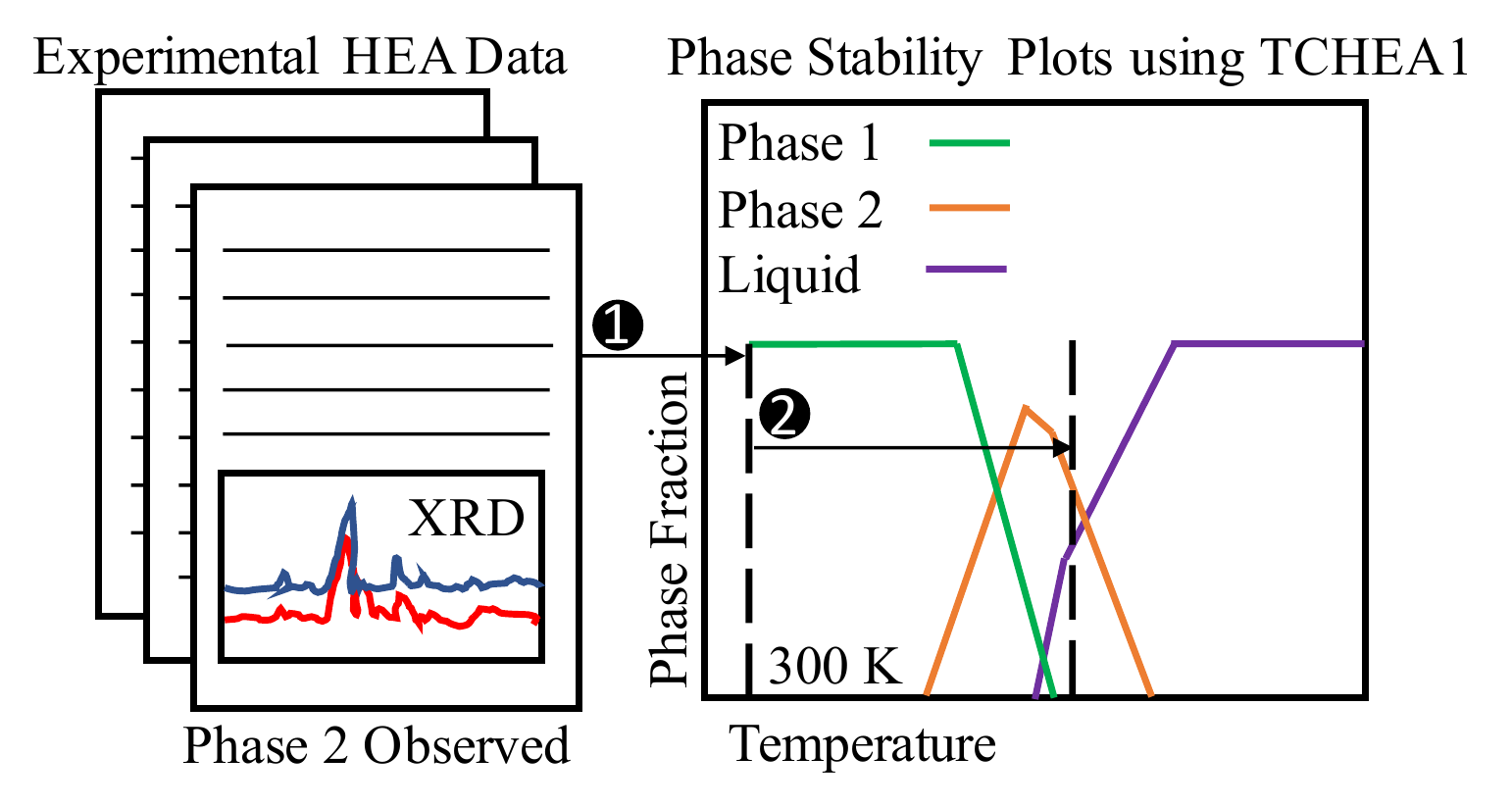}
\par\end{centering}
\caption{A systematic verification of Toda-Caraballo~\cite{toda2016criterion} dataset.}
\label{fig:verification}
\end{figure}

In this work, we carried an extensive evaluation of TCHEA1 using the dataset examined by Toda-Caraballo \etal~\cite{toda2016criterion}. We collected the appropriate literature as listed in the dataset and corrected for evident inconsistencies. For example, only a single FCC solution was listed by Toda-Caraballo \etal~\cite{toda2016criterion} for the AlCrCuFeNi$_2$ alloy while duplex microstructures (FCC+BCC) had been experimentally observed~\cite{guo2011effect}. A full report on the detailed verification of the Toda-Caraballo data set is included as Supplementary Material.

We examined each of the articles reported and tried to compare the reported phase constitutions to those predicted using TCHEA1 (Fig. ~\ref{fig:verification}). One thing of note is the fact that the vast majority of the alloys reported were characterized in the as-cast conditions and thus do not really even approach an equilibrium state. The AlCoCr$_{1.5}$FeMo$_{0.5}$Ni alloy, whose microstructure in the as-cast condition is reported to consist of $B2+\sigma$~\cite{hsu2011superior}, while phase stability predictions (see \textbf{Supplementary Material}, Fig. E1) indicate that a disordered BCC solid solution exists over a small range of temperatures below the solidus and that then decomposes into the $\sigma$ phase. In our analysis, this particular alloy was considered to be a good match between experiments and predictions. Another example is the case of the as-cast AlCoCrFeNiTi$_{1.5}$ system, with a reported phase constitution of two BCC phases as well as a Laves phase, as determined by XRD phase analysis~\cite{zhou2007solid}. Our phase stability predictions indeed show three phases upon solidification of the alloy (see \textbf{Supplementary Material}, Fig. E2), identified as an ordered B2, a disordered BCC as well as a Laves (C14) phase. At lower temperatures ($<$ 1000 K) we predict the formation of a Heusler (L2$_1$) phase as a result of further ordering of B2, although this phase is not reported~\cite{zhou2007solid}. Accounting for possible kinetics sluggishness as well as difficulties in detecting characteristic peaks of an L2$_1$ structure in a B2-ordered matrix we considered this case once again as a match between observations and predictions.

\begin{table}[]
\centering
\caption{Overview of comparison between Toda-Caraballo dataset and phase stability predictions with TCHEA1 database.}
\label{tab:comparison}
\begin{tabular}{llllllllll}
\cline{1-2}
Alloys Studied                          & 216 &  &  &  &  &  &  &  &  \\
Alloys Matching TCHEA1                  & 153 &  &  &  &  &  &  &  &  \\ \cline{1-2}
Alloys in As-Cast State                 & 180 &  &  &  &  &  &  &  &  \\
Alloys in As-Cast State Matching TCHEA1 & 127 &  &  &  &  &  &  &  &  \\ \cline{1-2}
Single Phase BCC Alloys                 & 23  &  &  &  &  &  &  &  &  \\
Single Phase BCC Alloys Matching TCHEA1 & 20  &  &  &  &  &  &  &  &  \\ \cline{1-2}
Single Phase FCC Alloys                 & 26  &  &  &  &  &  &  &  &  \\
Single Phase FCC Alloys Matching TCHEA1 & 22  &  &  &  &  &  &  &  &  \\ \cline{1-2}
\end{tabular}
\end{table}

Table~\ref{tab:comparison} summarizes the comparisons made with the dataset and a more detailed description of the comparisons is available in the \textbf{Supplementary Material}. Overall, we estimate that in 70.8\% of the reported 216 alloys described in Toda-Caraballo\etal~\cite{toda2016criterion} there was good agreement between the computed phase stabilities and experimental observations. The comparison accounted for possible kinetic factors and this was necessary as only 17\% of all alloys considered underwent long-term (relatively speaking) annealing treatments. Further improvement of the TCHEA1 database is necessary~\cite{chen2017database}, possibly by adding DFT-based alloy energetics~\cite{lederer2017search} as well as by updating the models through newly acquired experimental observations. In this work, we considered that the quality of the thermodynamic descriptions contained in TCHEA1 was sufficient to carry out the systematic exploration of the HEA alloy space.

\section{Exploration of the HEA Space through the Constraint Satisfaction Algorithm}
In this work, we focus on the CoCrFeMnNi system, a prototypical HEA, in order to demonstrate the ability of the CSA to identify targeted phase stability regions in the multi-dimensional HEA space. The solution to the CCSP in the context of materials design consists of defining (non-linear) constraints that together define a desired phase constitution state. The CSA is then deployed to identify all the regions in the (C-T)We se space that satisfy the constraints. As an example, one could define a constraint as the requirement that a given region in the C-T space corresponds to a single phase field within a specified temperature range, with at most $\chi$\% volume fraction of secondary phases, with the primary phase being either FCC, BCC or HCP. More sophisticated constraints can be defined, such as the requirement that an alloy that exists as a single-phase solid solution within a temperature range, undergoes the precipitation of a single (possibly strengthening) secondary phase at lower temperatures, etc.

The search for single-phase solid solutions was defined in terms of constraints as follows: single-phase solid solutions were sought in which a given composition deviated about +/- 5\% from the pure equiatomic composition. For example, in a quaternary system, each constituent was allowed to vary between 20 and 30 at.\%. The C-T space was explored within the 500-2000 K range. We would like to note that the constraints fed into the CSA \emph{can be relaxed well beyond the near-stoichiometry} used in this work to validate this exploration framework of the HEA space. For example, without loss of generality, one could set the composition ranges for each of the constituent of the alloy to be within the 5-40 \%––further exploration of the phase stability space in some select HEA systems will be subject of future work.

To account for the fact that some of the target regions in the solution space could correspond to ordered phases, we defined solid solutions as regions consisting of at least 99\% of BCC, FCC, HCP or their ordered variants. In this work, the degree of ordering of a given phase was determined by examining the site occupancy of a given phase after Gibbs energy minimization, as suggested in ~\cite{chen2017database}.

\subsection{The Search for Ternary Single-Phase Solid Solutions}
The CoCrFeMnNi has ten possible ternaries, CoCrFe, CoCrMn, CoCrNi, CoFeMn, CoFeNi, CoMnNi, CrFeMn, CrFeNi, CrMnNi and FeMnNi and we searched those spaces for single-phase solid solutions in which the composition of the constituents was varied within the 28-38 \% range. The CSA implementation used a Genetic Algorithm~\cite{galvan2017constraint}
to expand the SVDD boundaries and for all cases we ran the CSA for 75 generations with 75 individuals belonging to each generation. Here individuals of a population corresponded to a single Gibbs energy minimization using the TCHEA1 database.

\begin{table*}[hbt!]
\centering
\caption{Single-phase solid solutions in ternary alloys in the CoCrFeMnNi quinary system.}
\label{tab:ternary}
\begin{tabular}{lllll}
\hline
\multirow{2}{*}{Ternary} & \multirow{2}{*}{CSA SS} & \multirow{2}{*}{\begin{tabular}[c]{@{}l@{}}Temperature\\ Range\end{tabular}} & \multirow{2}{*}{Experiments~\cite{wu2014recovery}} & \multirow{2}{*}{\begin{tabular}[c]{@{}l@{}}Homogenization \\ Temperature  (K)\end{tabular}} \\
                         &                         &                                                                              &                              &                                                                                             \\ \hline
CoCrFe                   & BCC,FCC                 & 1572-1695,1178-1681                                                          & N/A                          & N/A                                                                                         \\
CoCrMn                   & BCC, HCP                & 1107-1540, 645-1034                                                          & Multi                        & 1373                                                                                        \\
CoCrNi                   & FCC                     & 819-1716                                                                     & FCC                          & 1473                                                                                        \\
CoFeMn                   & FCC                     & 745-1581                                                                     & Multi                        & 1473                                                                                        \\
CoFeNi                   & FCC                     & 880-1731                                                                     & FCC                          & 1473                                                                                        \\
CoMnNi                   & L1$_2$                  & 688-1487                                                                     & FCC                          & 1373                                                                                        \\
CrFeMn                   & BCC                     & 1324-1713                                                                    & N/A                          & N/A                                                                                         \\
CrFeNi                   & FCC                     & 1044-1672                                                                    & FCC                          & 1473                                                                                        \\
FeMnNi                   & L1$_2$                  & 542-1509                                                                     & FCC                          & 1373                                                                                        \\
CrMnNi                   & L1$_2$                  & None                                                                         & Multi                        & 1323                                                                                        \\ \hline
\end{tabular}
\end{table*}

Table ~\ref{tab:ternary} shows that a single-phase solid solution was found in the all the ternaries, except for CrMnNi, in which the CSA failed to identify a single-phase solid solution over the entire temperature range considered. Upon further examination, however, we identified a relatively wide temperature range (800-1400 K) in which at least 75\% of the microstructure should have been constituted by an L1$_2$ phase (See Fig. A3 in the \textbf{Supplementary Material}). Besides the work of Bracq	 \etal~\cite{bracq2017fcc} on the CoCrFeMnNi HEA, there are not many attempts in the literature to experimentally determine the compositional and temperature range of a single-phase solid solution in an HEA. Such works on the ternaries and quaternaries that comprise the CoCrFeMnNi alloy would be ideal for a thorough comparison with the results derived from the CSA-based search. 

In this study, the results are compared against phase analysis based on XRD on equiatomic compositions available in the literature~\cite{wu2014recovery}. Qualitatively, the phase constitution observations available seem to match what is predicted through the CSA. An exception would be CoFeMn which is reported to consist of multiple phases over a wide temperature range, while the CSA reports a finite temperature range in which a single-phase FCC solid solution is stable. This could be because the disordered FCC phase is not completely stable at the characterization temperature~\cite{wu2014recovery}. 

Both CoMnNi and FeMnNi are predicted to have a single-phase L1$_2$ structure through the CSA, but the equiatomic compositions are characterized experimentally as consisting of a single-phase FCC solid solution. An L1$_2$ designation implies an ordered structure although closer inspection of the predicted site fractions as a function of temperature suggests that concentration differences among sublattices in the underlying FCC lattice is within 0.1 (see \textbf{Supplementary Material} Fig. F1 and F2). Only very sophisticated analysis of the XRD signal could detect such subtle differences in site occupancy and thus it is likely that predictions truly reflect the thermodynamic behavior in these systems.

\begin{figure}
\begin{centering}
\includegraphics[width=0.9\columnwidth]{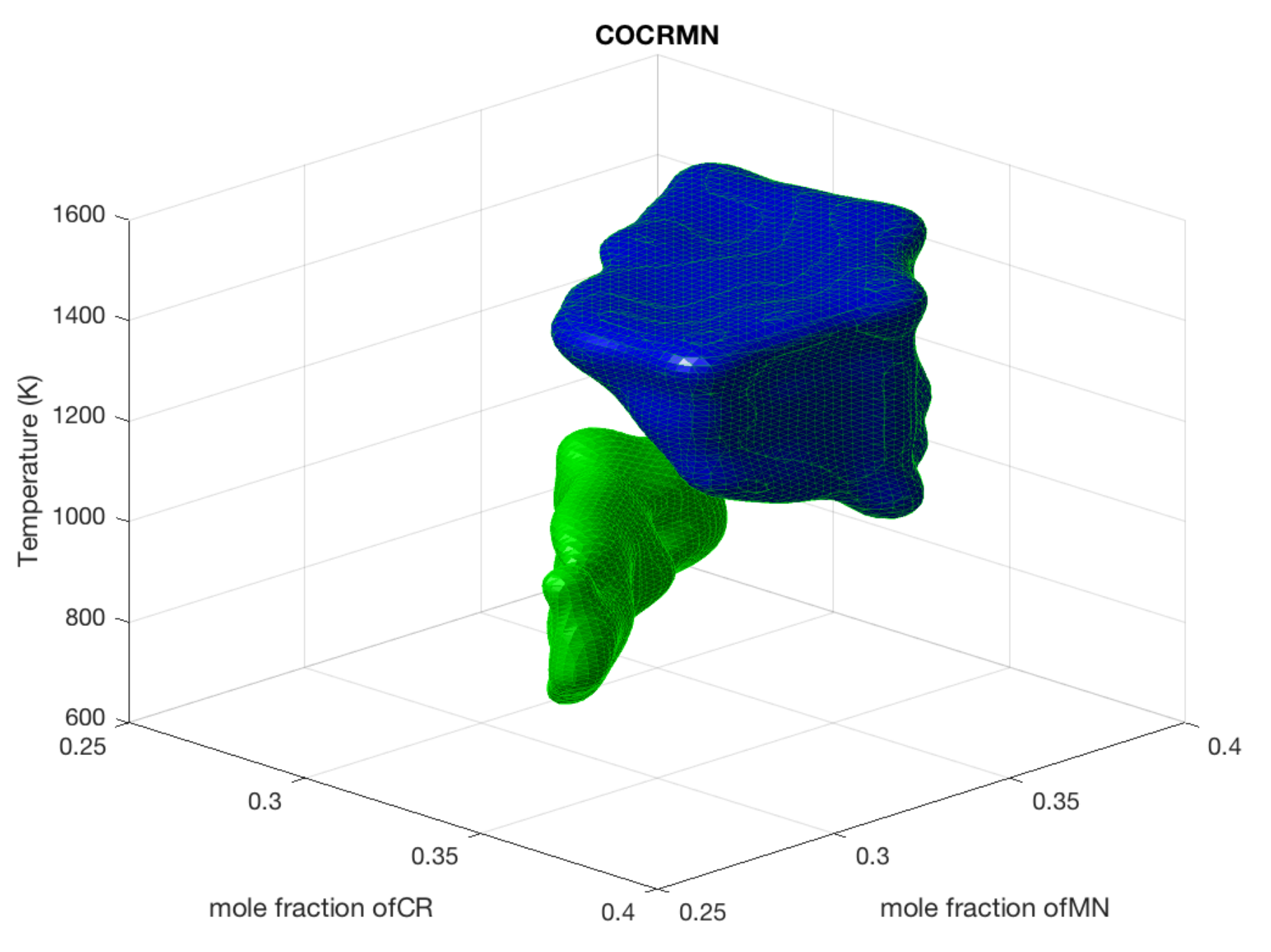}
\par\end{centering}
\caption{Constraint Satisfaction Algorithm (CSA) prediction of single phase BCC and HCP solid solution ranges of stability. BCC single-phase solid solution is depicted in blue, while HCP single-phase solid solution is depicted in green (color online). Composition of the constituents was allowed to vary within the 28-38 \% range. }
\label{fig:CoCrMn}
\end{figure}

In the case of a ternary system, visualization of the SVDD representation of the solution to the CSA can be easily realized. An example of such visualization is shown in Fig.~\ref{fig:CoCrMn} which depicts the regions of stability of single-phase BCC and HCP solid solutions in the CoCrMn ternary system. As stated above, composition of the three constituents was allowed to vary within the 28-38 \% range. The figure shows clearly that this system switches between a BCC to an HCP-dominated single-phase solid solution "island of stability" as temperature decreases. This is consistent with the phase stability behavior of cobalt and it seems to suggest that the phase stability of this ternary is dominated by the behavior of cobalt. Visualization of the other ternaries is available in the \textbf{Supplementary Material}, Figs. B1-B8.

\subsection{The Search for Quaternary Single-Phase Solid Solutions}
The CoCrFeMnNi system has five possible quaternaries that can form from the five elements: CoCrFeMn, CoCrFeNi, CoCrMnNi, CoFeMnNi, and CrFeMnNi. A purely equiatomic concentration would result in each element contributing to 25\% of the composition. The domain of the composition search space for each element as set within the 20-30\% range. This is a four-dimensional problem, with three of the dimensions being concentrations and the last being temperature. In the CSA, we defined the maximum number of generations to 150, with a population of 150 individuals per generation.

\begin{table*}[]
\centering
\caption{Single-phase solid solutions in quaternary alloys in the CoCrFeMnNi quinary system.}
\label{tab:quaternary}
\begin{tabular}{lllll}
\hline
\multirow{2}{*}{Quaternary} & \multirow{2}{*}{CSA SS} & \multirow{2}{*}{\begin{tabular}[c]{@{}l@{}}Temperature\\ Range\end{tabular}} & \multirow{2}{*}{Experiments~\cite{wu2014recovery}} & \multirow{2}{*}{\begin{tabular}[c]{@{}l@{}}Homogenization \\ Temperature  (K)\end{tabular}} \\
                            &                         &                                                                              &                              &                                                                                             \\ \hline
CoCrFeMn                    & BCC,FCC                 & 1325-1606,986-1585                                                           & Multi                        & 1373                                                                                        \\
CoCrFeNi                    & FCC                     & 827-1726                                                                     & FCC                          & 1473                                                                                        \\
CoCrMnNi                    & FCC*,L1$_2$             & 1448-1474,677-1548                                                           & FCC                          & 1373                                                                                       \\
CoFeMnNi                    & L1$_2$                  & 766-1573                                                                     & FCC                          & 1373                                                                                        \\
CrFeMnNi                    & L1$_2$                  & 1009-1548                                                                    & Multi                        & 1373                                                                                        \\ \hline
\end{tabular}
\end{table*}

Table ~\ref{tab:quaternary} shows the predicted and experimentally-determined phase stability in the quaternary systems. When compared to XRD phase observations of equiatomic compositions, the CSA shows good qualitative agreement. Our predictions suggest that the FCC single-phase region in CoCrMnNi (denoted by an asterisk) is negligible compared to the L1$_2$ region.  The CrFeMnNi single L1$_2$ phase at higher temperatures may not be stable at lower temperatures, leading to a multiphase structure during characterization––via XRD. The CoCrMnNi and CoFeMnNi compositions largely showed an L1$_2$ structure through the CSA, while experiments report a single-phase FCC solid solution via XRD phase analysis~\cite{wu2014recovery}. Analysis of the site fractions of these equiatomic compositions, however, showed very small deviations from equal partitioning among the different sublattices in the FCC sublattice (see \textbf{Supplementary Material}, Figs. F3-F5). Such a small and subtle differences in site occupancy would be challenging to detect experimentally and thus we consider that the CSA-based calculations and experiments agree at least within the limits of resolution of the experiments used to examine the phase constitution in these systems. 

\begin{figure*}[hbt!]
\begin{centering}
\includegraphics[width=0.9\textwidth]{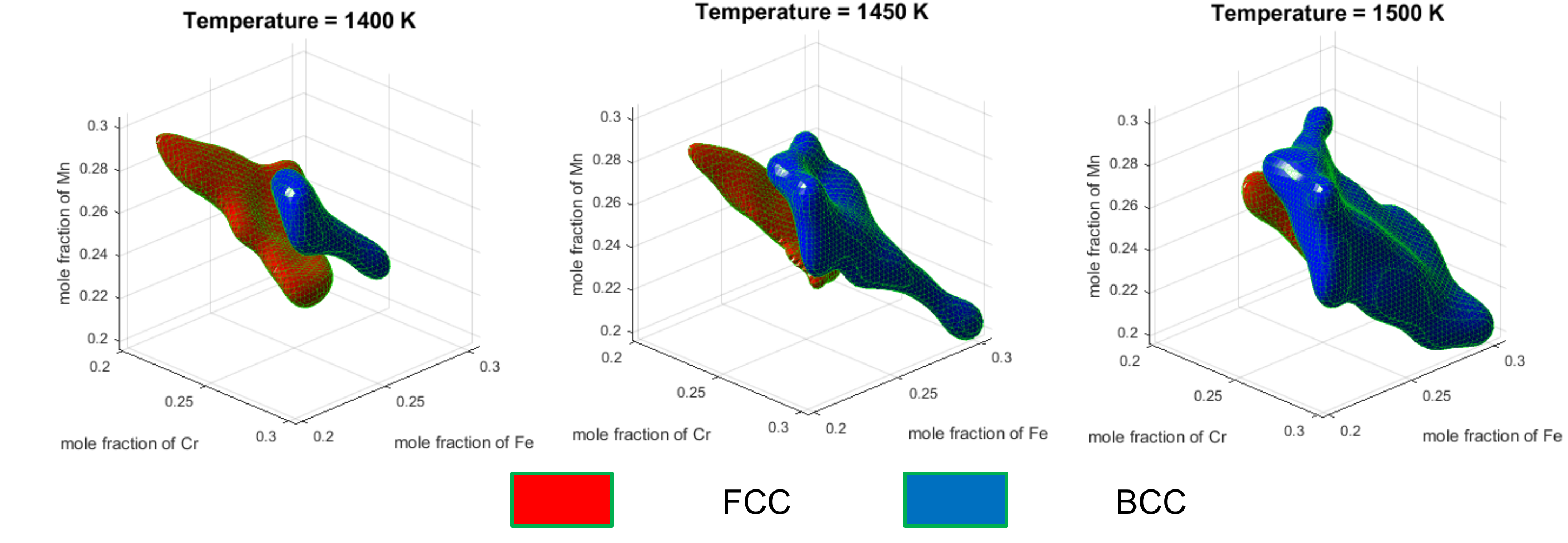}
\par\end{centering}
\caption{Constraint Satisfaction Algorithm (CSA) prediction of single phase BCC and FCC solid solution ranges of stability. BCC single-phase solid solution is depicted in blue, while FCC single-phase solid solution is depicted in red (color online). Composition of the constituents was allowed to vary within the 20-30 \% range.}
\label{fig:CoCrFeMn}
\end{figure*}

Visualizing the SVDD phase stability boundaries derived from the CSA in quaternary systems is challenging because of the extra dimension. Yet, to have a visual representation of the so-called "stability islands" in these higher order systems it is possible to make projections of this 4-D space in a 3-D surface. This corresponds to visualizing the stability of the system at different temperatures. Fig. ~\ref{fig:CoCrFeMn} shows the stability of FCC and BCC single-phase solid solutions. The figure shows clearly how the BCC phase becomes increasingly stable as temperature increases. This is in line with the fact that BCC lattices tend to have higher entropy than FCC ones~\cite{friedel1974stability}. These results also put into question the use of the Valence Electron Cocentration (VEC) as a marker for stability of BCC vs FCC HEA systems since in CoCrFeMn (under constant VEC) the two phases switch their relative stability with temperature~\cite{guo2011effect}. Visualizations for the remaining quaternary systems are included in the \textbf{Supplementary Material}, Figs. C1-C4. A notable result is that in the remaining systems, the central region of the composition space is dominated by FCC single-phase solid solutions over the entire temperature range considered.

\subsection{The Search for Quinary Single-Phase Solid Solutions}
The CoCrFeMnNi system has an equiatomic composition when each of its elements are at 20\% concentration. As per our specification of the allowable search space for single-phase solid solutions, we restricted the composition of each constituent within the 15-25 \% range. In this case, the CSA must identify suitable regions that satisfy the constraints imposed over a five-dimensional space––four constituents, plus temperature. Since the CSA does not have a formal termination criterion, we limited the search over this five-dimensional space to 200 generations with 200 individuals generated in every generation.

\begin{figure*}[hbt!]
\begin{centering}
\includegraphics[width=0.9\textwidth]{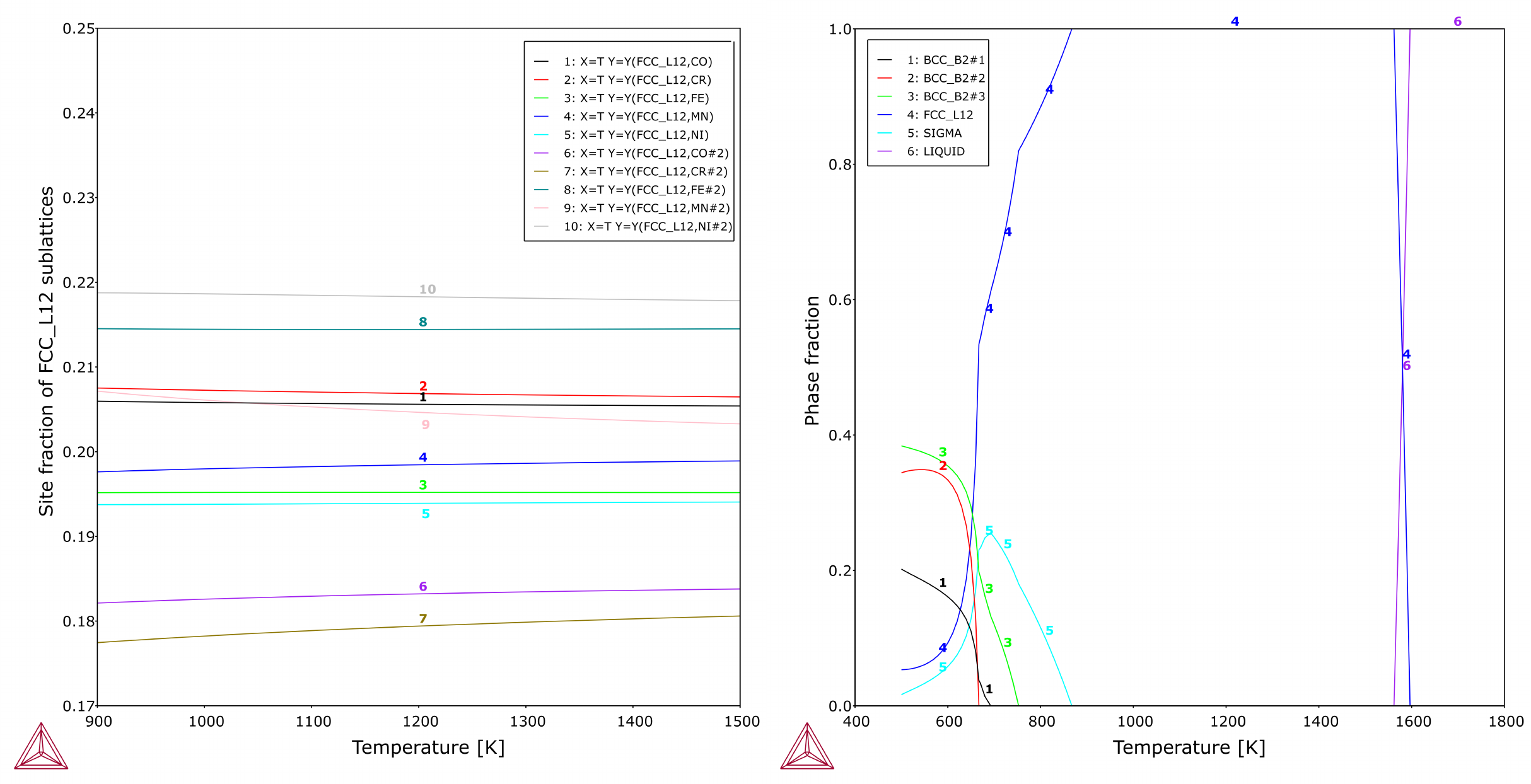}
\par\end{centering}
\caption{(left) computed evolution of sublattice site fraction in the L1$_2$ phase of the equiatomic FeCoCrMnNi system as a function of temperature. (right) computed phase stability in the equiatomic FeCoCrMnNi system as a function of temperature.}
\label{fig:CoCrFeMnNia}
\end{figure*}

The main single-phase solid solution region found in the CoCrFeMnNi alloy corresponds to the L1$_2$ phase within the 751-1601 K temperature range. This temperature range of stability is similar to the one determined by Bracq \etal~\cite{bracq2017fcc} at near-equiatomic concentrations of the five component HEA. The computed stability search for the disordered FCC region yielded a very small stability range for this phase, which is surprising because the CoCrFeMnNi composition is commonly cited as having an FCC phase~\cite{bracq2017fcc,wu2014recovery}. We examined the calculations further by computing the evolution of site fractions with temperature as shown in Fig.~\ref{fig:CoCrFeMnNia}, which shows that the elemental site fractions in the two sublattices used to describe this phase are within 0.02, indicating that this is essentially a disordered phase. This occupation degeneracy holds even at relatively low temperatures (1000 K). The calculated phase stability in the CoCrFeMnNi equiatomic system is also shown in Fig.~\ref{fig:CoCrFeMnNia}. 

\begin{figure*}[hbt!]
\begin{centering}
\includegraphics[width=0.9\textwidth]{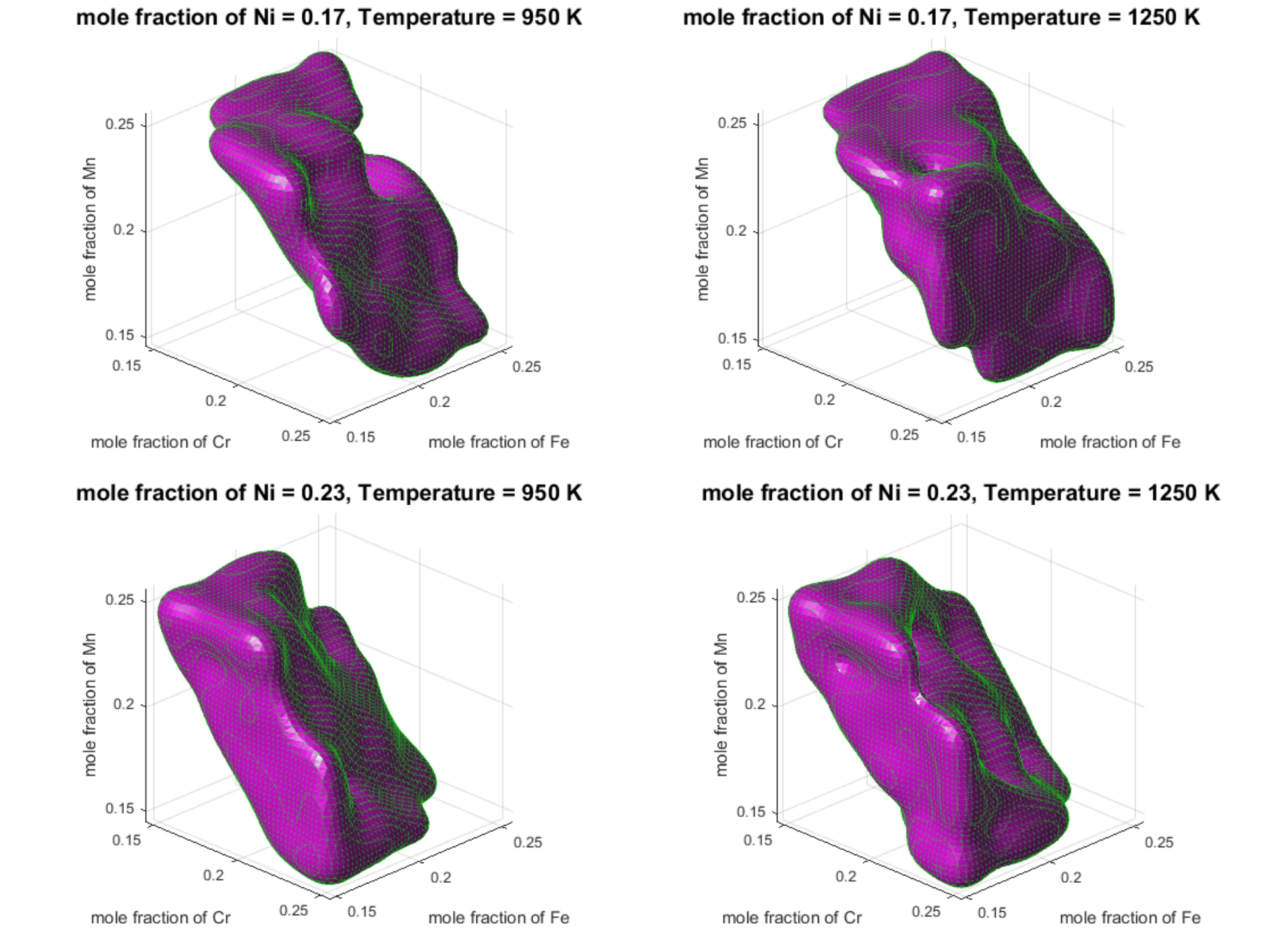}
\par\end{centering}
\caption{Evolution of FCC phase stability in the CoCrFeMnNi quinary system with changes in temperature and Ni concentration.}
\label{fig:CoCrFeMnNib}
\end{figure*}

The visualization for the change in the stability range of the single-phase FCC solid solution as a function of temperature and Ni concentration in the CoCrFeMnNi system can be seen in Fig.~\ref{fig:CoCrFeMnNib}. The figure compares two Ni concentrations (17 and 23 \%) and two temperatures (950 and 1250 K). The figure shows that the extent of stability of the FCC phase in the central region of the quinary composition space does not change significantly with composition or temperature, although the results from the CSA-based exploration indicate that the single-phase FCC solid solution stability range increases with temperature (at lower temperatures this phase competes for stability with the $\sigma$ phase, for example) and Ni content.

\subsection{Evaluating the Performance of the CSA}

\begin{figure*}[hbt!]
\begin{centering}
\includegraphics[width=0.9\textwidth]{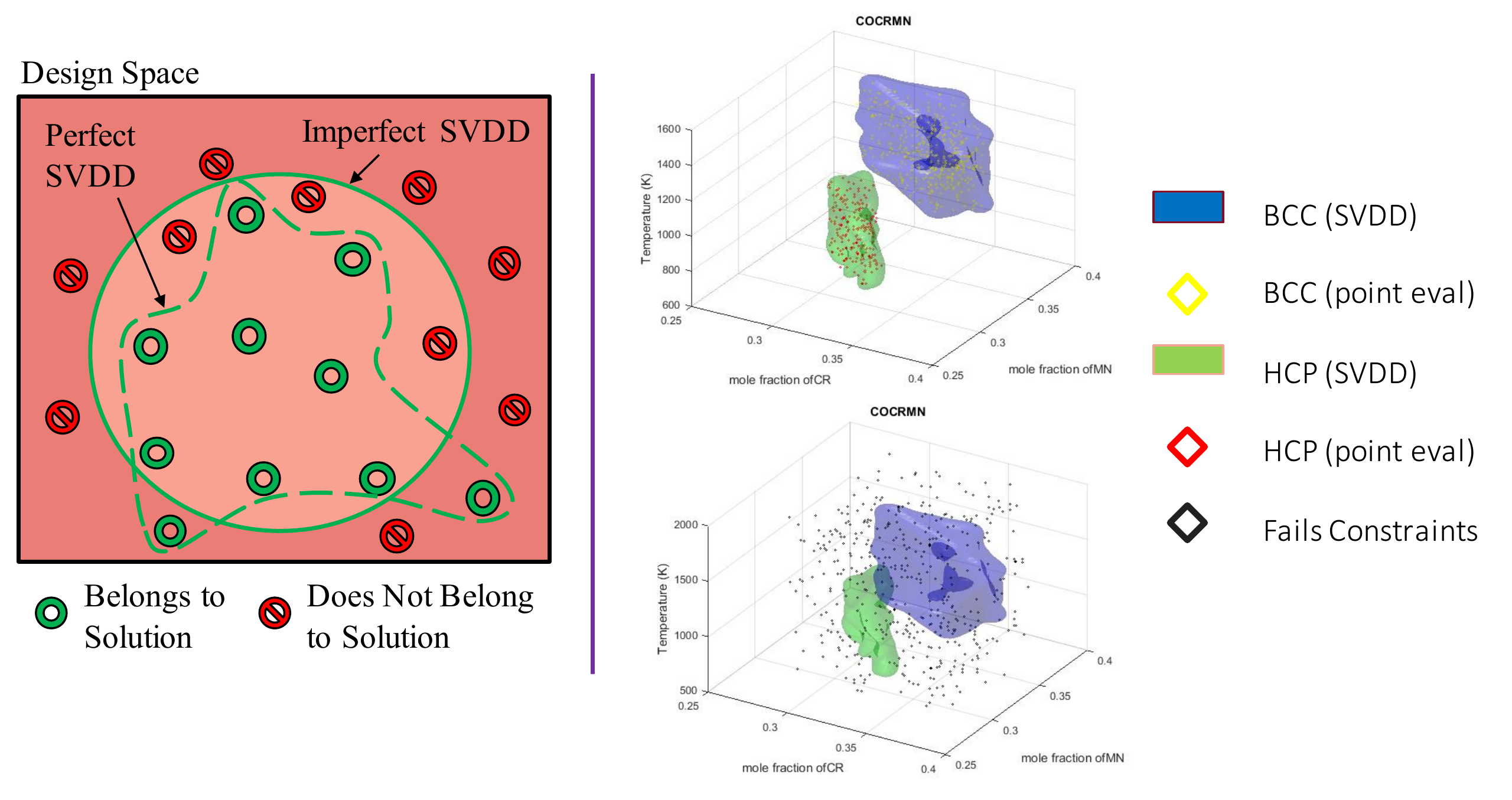}
\par\end{centering}
\caption{An example of the evaluation of the SVDD classifier. (left) a perfect SVDD is compared to an imperfect one: an SVDD with high precision is one in which the solution boundary encompasses the entirety of the target; while one with high recall is one in which the entirety of the target has been captured by the SVDD. (right) comparison between the SVDD determined by the CSA and actual phase stability calculations in the ternary CoCrMn system: only a small fraction of the points indicated as belonging to the SVDD fail to meet the constraints.}
\label{fig:SVDD}
\end{figure*}

To quantify the reliability of the CSA up to five dimensions, the precision, recall, and misclassification rates––see Fig.~\ref{fig:SVDD} (left)––were calculated for one ternary, one quaternary, and the quinary systems. \emph{Precision} represents the percentage of data in the solution boundary (as identified by the SVDD) that is part of the target data. \emph{Recall} represents the percentage of the target data that is in the solution boundary. \emph{Misclassification}, on the other hand, corresponds to the percentage of the data that is incorrectly classified (i.e. falst positives and false negatives relative to the totality of the data). The method of calculating these metrics was adapted from Galvan \etal~\cite{galvan2017constraint}, using 10$^6$ random points within the temperature and composition ranges of the respective alloy and calculating their phase stability in Thermo-Calc. The points that satisfy the constraints as defined previously correspond to the target data. Evaluation of the SVDDs were done directly over this synthetic dataset. An example of the evaluation of the performance of the CSA during the exploration of the CoCrMn can be found in Fig.~\ref{fig:SVDD} (right), which shows how only a very small fraction of the points in the space actually fail the constraints.
 
\begin{table}[]
\centering
\caption{Precision, recall and misclassification rate of the SVDDs determined by the CSA for three multi-component systems in the CoCrFeMnNi quinary system.}
\label{tab:svdd}
\begin{tabular}{llll}
\hline
                                & Precision & Recall  & Missclass. Rate \\ \cline{2-4} 
\multicolumn{1}{l|}{CoCrMn}     & 94.68\%   & 90.87\% & 1.94\%          \\
\multicolumn{1}{l|}{CoCrFeMn}   & 93.5\%    & 72.39\% & 1.03\%          \\
\multicolumn{1}{l|}{CoCrFeMnNi} & 83.64\%   & 97.69\% & 5.02\%          \\ \hline
\end{tabular}
\end{table}

The precision, recall and misclassification rates of the SVDDs generated by the CSA for the CoCrMn, CoCrFeMn and CoCrFeMnNi systems are shown in Table~\ref{tab:svdd}. Overall, the results are quite satisfactory as in general the SVDDs for all the systems had relatively high rates of precision and recall, with the precision rate being slightly lower for the quinary than for the ternary and quaternary systems. The recall rate, on the other hand, exhibited a somewhat unusual trend as the system of intermediate dimensionality (CoCrFeMn) showed the lowest recall. This is probably due to issues related with the specific characteristics of the system and the sampling, as the lower recall rate means that the SVDD was not able to capture all the points deemed to satisfy the constraints. Running the CSA for a higher number of generations (or iterations id boundaries are expanded through the bisection method) will tend to increase recall rates, as already demonstrated in our prior work~\cite{galvan2017constraint}. Table~\ref{tab:svdd} shows that for all the cases studied the SVDD had a relatively low misclassification rate, which implies a rather effective classifier.

 A further meassure of the effectiveness of the SVDD could be attained by comparing the computational cost of a targeted search enabled by the CSA with a grid-search as done previously in so-called high-throughput CALPHAD exploration of the HEA space~\cite{senkov2015acceleratedb,senkov2015accelerated}. We performed this evaluation for the CrFeNi by changing the concentration of each element by 0.5\% from 0-100\% and temperature intervals of 25K over a 300-1100 K range. A grid search took several days, while the exploration of the ternary systems took less than an hour to complete. Moreover, the CSA automatically encloses areas that satisfy the design criteria, without having to carry out further statistical analysis of the phase stability calculations. Based on its performance as a classification scheme and the efficiency with which it explores the multi-component space, the present CSA  is clearly a much more efficient way to search through the design space. Even better performance of the CSA can be achieved by expanding the boundaries of the SVDD through the bi-section method, rather than the Genetic Algorithm used in this work. This has been discussed in detail by Galvan \etal~\cite{galvan2017constraint}.

\section{Towards Microstructural Complexity: Finding Precipitation Strengthened HEAs}

\begin{figure}[hbt!]
\begin{centering}
\includegraphics[width=0.9\columnwidth]{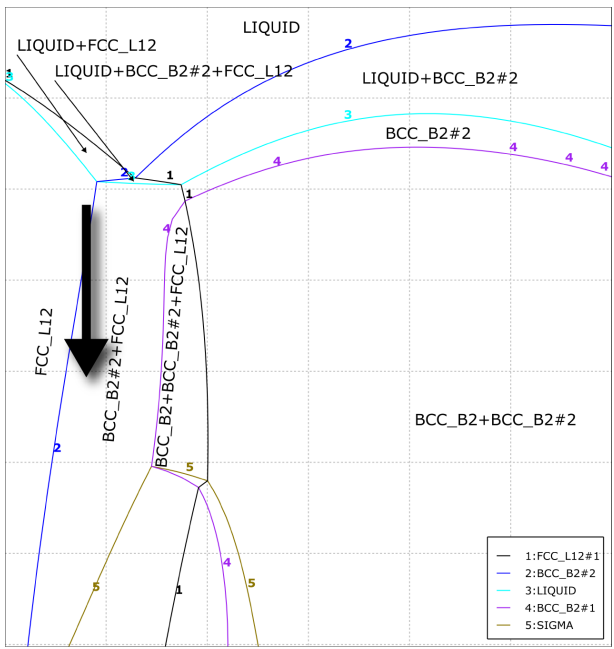}
\par\end{centering}
\caption{Al-CoCrFeNi isopleth showing a two-phase region under a single-phase solid solution region.}
\label{fig:AlCoCrFeNi}
\end{figure}

Over the past years, the focus on discovering single-phase solid solutions in the HEA space has slowly given way to the exploration of compositionally~\cite{miracle2017high} and microstructurally complex~\cite{miracle2017critical} systems. This emphasis has arisen naturally and is a common pattern encountered historically in all conventional structural alloy systems: microstructural and compositional complexity originate from the need to improve materials performance beyond what a single phase can achieve. Given the vastness of the HEA space, however, it is unrealistic that fully random explorations of the composition/microstructural space are likely to yield meaningful results anytime soon. Unfortunately, none of the existing computational and computer-aided approaches to the "design" of HEAs is capable of such a complex search.

In this work, we present, for the first time, the targeted CALPHAD-based search of microstructurally complex HEAs. As an example, our target was to identify a region in a multi-component space that was likely to yield a two-phase microstructure in which a minority phase could act as a strengthening phase. A prototypical HEA system is that of Al$_x$CoCrFeNi, 
whose microstructure has been observed to consists of an FCC matrix with B2 second phase precipitates over specific amounts of Al~\cite{kao2009microstructure}––Fig. ~\ref{fig:AlCoCrFeNi} shows the isopleth in the Al-CoCrFeNi system as calculated using the TCHEA1 database.

Using the CSA, the temperature and composition space where this system can be precipitation-hardened was searched. From an alloy design perspective the constraints in this case were established as follows:
\begin{itemize}
\item{Each of the elements was allowed to change its composition between the 10-30\% range.}
\item{The CSA then searched over the 1500-1800 K range to ensure a single-phase solid solution region. This is essential since the control of the precipitation process can only be achieved if second phase particles formed during initial synthesis of the material can be 'erased' from the microstructure.}
\item{If a given composition with the solid-solution window as specified above was found, the CSA would attempt to find, over a limited temperature range, a two-phase region. Such a second phase would have to be stable above 500 K as otherwise sluggish kinetics would prevent its precipitation.}
\item{If a two-phase region was found, the two phases must satisfy the following criteria. Firstly, the two phases must not result from spinodal decomposition. Secondly, if the primary phase is an ordered phase (B2, L1$_2$), then the secondary phase cannot be its disordered counterpart (BCC, FCC, respectively).}
\item{If a composition passed these criteria, it is considered to have the potential to result in a precipitation-hardanable alloy.}
\end{itemize}

\begin{figure*}[hbt!]
\begin{centering}
\includegraphics[width=0.9\textwidth]{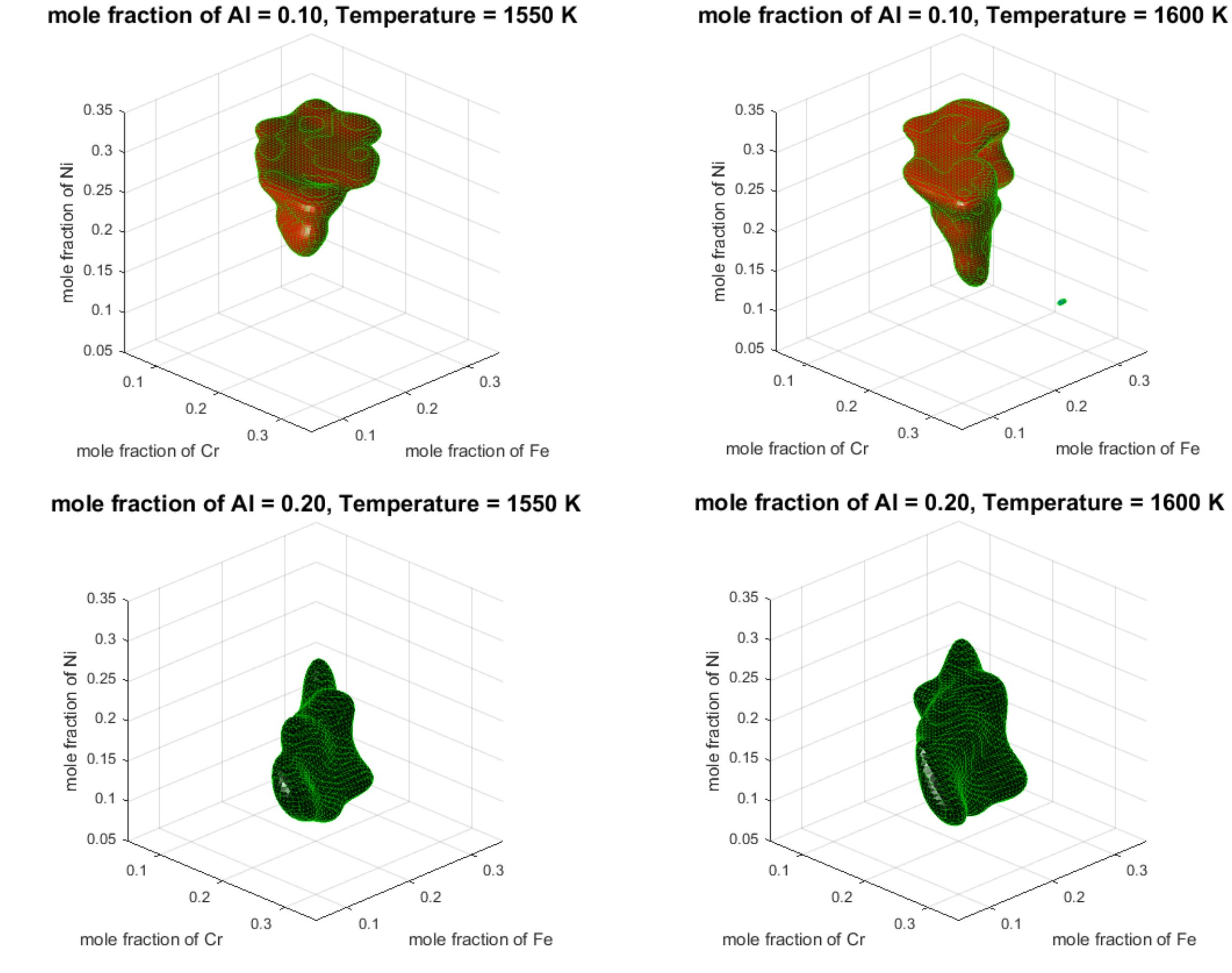}
\par\end{centering}
\caption{Stability of regions capable of precipitation strengthening in the AlCoCrFeNi system for different Al concentrations. Regions with FCC-based matrix is depicted as red, while BCC-based alloys are depicted as black.}
\label{fig:precipitate}
\end{figure*}

To solve this five-dimensional problem, we restricted the search to 200 generations with 200 individuals per generation.The search for the potential precipitate hardening regions in the AlCoCrFeNi system yielded a small region of BCC (within 1579 – 1642 K) and large regions of both FCC and B2 phases (within 1500 – 1654 K and 1500 – 1707 K, respectively) that could be precipitation-hardened as per the constraints described above. Fig.~\ref{fig:precipitate} depicts the stability regions for alloys in the AlCoCrFeNi system likely to result in precipitation-strengthened microstructures as prescribed above. As seen in Fig.~\ref{fig:precipitate}, a lower atomic concentration of Al results in having an FCC (red) region ready to transit to a two-phase region, while higher Al concentrations result in having a BCC/B2 (black) region ready to transit to a two-phase region.

\section{Beyond Phase Stability}

\begin{figure}[hbt!]
\begin{centering}
\includegraphics[width=0.9\columnwidth]{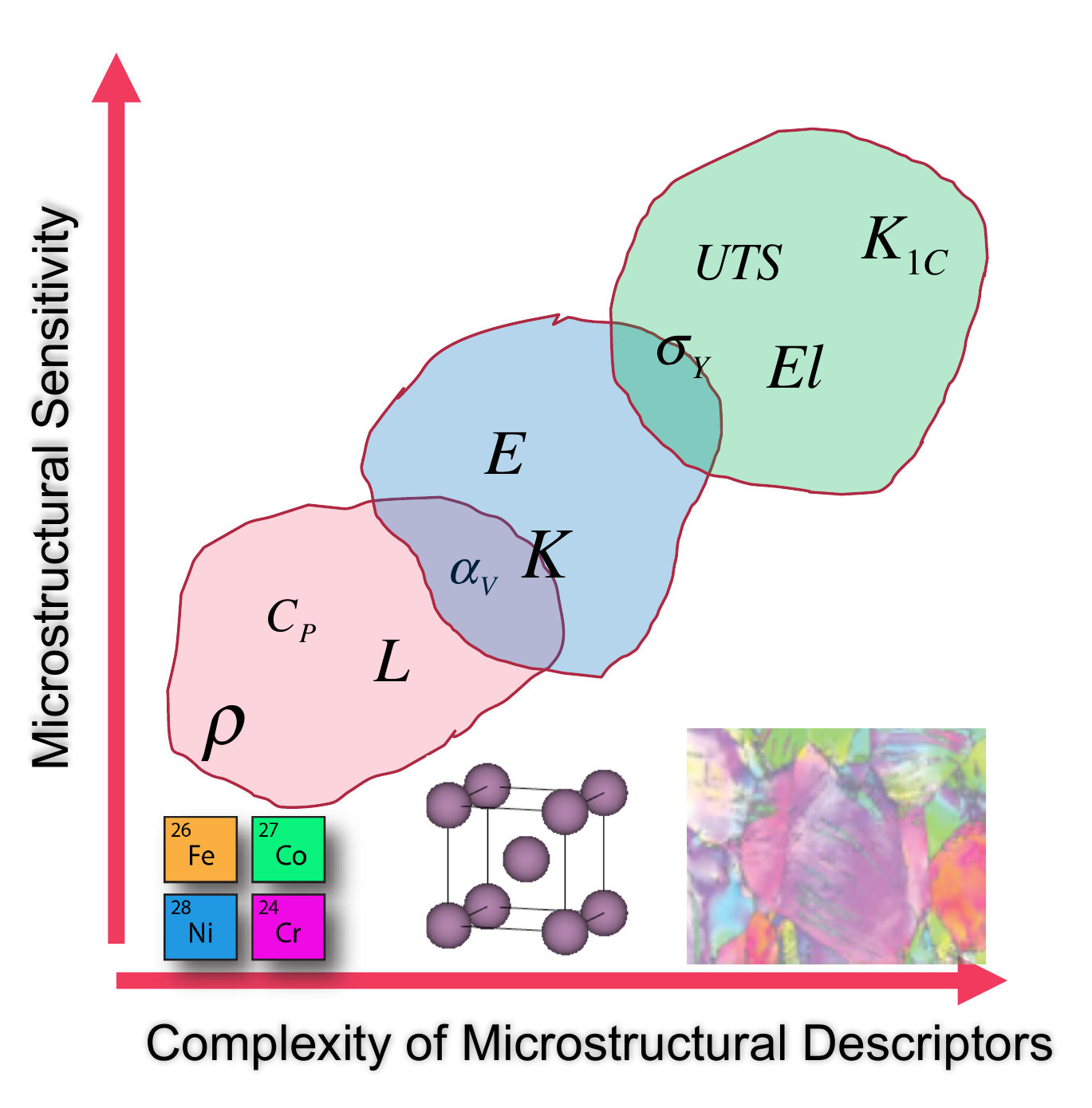}
\par\end{centering}
\caption{Hierarchy of (mechanical/thermodynamic) properties~\cite{toda2013unravelling} as a function of complexity in microstructural descriptors: density ($\rho$), melting point ($T_m$), latent heat of fusion ($L$), specific heat ($C_P$), thermal expansion coefficient ($\alpha_V$), thermal conductivity ($K$), Young's modulus (E), yield strength ($\sigma_Y$), elongation ($El$), fracture toughness ($K_{1C}$).}
\label{fig:properties}
\end{figure}

The proposed framework enables the efficient exploration of alloy design spaces but it is also important to note that the CSA does not operate exclusively on the alloy phase stability space. In fact, the proposed framework can incorporate arbitrary and non-linear constraints, as long as the the constraint(s) can be expressed in terms of either phase constitution or composition and their evaluation can be carried out using code that can be integrated within the CSA framework––this last requirement is extremely easy to fulfill as the CSA is implemented in MATLAB and constraint evaluations can be done both on and offline. 

As a concrete example of possible improvements upon the CSA-based search for candidate compositionally and microstructurally-complex formulations within the HEA space is the incorporation of constraints based on other properties of interest~\cite{gorsse2017mapping}. In fact, in a sense there is a hierarchy of structure-property connections based on the sensitivity of a given property of material performance metric to the complexity of the microstructural description one needs to establish such connections. 

Toda-Caraballo \etal~\cite{toda2013unravelling} for example, recently carried out an extensive statistical analysis of thermal and mechanical properties across a vast alloy space spanning the major families of structural alloys––at corners of the composition space––currently known. Their analysis tried to reduce the dimensionality of the problem through the use of Principal Component Analysis (PCA) and, based on the effectiveness of the PCA-based dimensional reduction they grouped the property superset into three distinct groups depending on their sensitivity to microstructure (see Fig.~\ref{fig:properties}). 

Group I (e.g. $\rho$, $T_m$), as defined by Toda-Caraballo \etal~\cite{toda2013unravelling} were properties in which the statistical models (as a function of composition) had the best success and corresponded to properties dominated by electronic and atomic-scale interactions (interatomic distance, bonding character, atomic weight). Group II were properties that depended on both chemistry as well as information on lattice structure, while Group III (e.g. $\sigma_y$, $K_{1C}$) were highly sensitive to microstructure and models based on chemistry alone are not likely to be effective. It is relatively straightforward to envision expansions to the current CSA-based approach to alloy discovery where constraints are written in terms of ranges of acceptable values for at least some properties---as suggested by Gorsse~\cite{gorsse2017mapping}---that are least sensitive to microstructure. 

Statistical models alone, however, are not sufficiently predictive~\cite{toda2013unravelling} and to alleviate this deficiency we could potentially make use of physics-based models. In the context of design of HEAs, for example, Toda-Caraballo \etal recently presented a predictive approach for the solid solution hardening (SSH) in HEAs. Their model was based on Labusch's statistical theory of solid solution hardening~\cite{labusch1970statistical}, modified using Gyben and Deruyttere's formalism~\cite{gypen1977multi} for concentrated alloys and Moreen's approach to predict the lattice parameters of solid solutions~\cite{moreen1971model}. Toda-Caraballo 's model is able to account for the contributions of lattice distortion and elastic misfit to predict accurately SSH in available HEAs. Such a model could be incorporated into our proposed CSA by requiring that a given solid solution has a minimum level of SSH allowable, for example.

In summary, it is evident from the examples given above that the exploration of the HEA space with the proposed CSA approach can go beyond the identification of regions in the HEA space with suitable phase constitutions. In fact, the framing of the problem in terms of (non-linear) constraints enables the expansion of the alloy search framework to incorporate a range of properties that can be used for further screening of potential HEAs as candidates for further (experimental) investigation.

\section{Conclusion}

In this work, we have presented a novel approach towards the targeted discovery of novel compositions in the HEA space, which is based on the solution to the inverse phase stability problem. We have described how this problem can be mapped to a so-called continuous constraint satisfaction problem and have presented a novel algorithm to solve it, the CSA. The CSA shows its ability to model the evolution of phase stability in multi-component systems such as HEAs. This approach can be used as a tool to discover and design alloys with tailored phases and properties. 

The CSA has also showcased its ability to determine alloy compositions that are susceptible to certain processing conditions, such as precipitation hardening. Examples of three-, four-, and five-dimensional searches with the CSA were tested to have relatively high precision and recall rates along with low misclassification rates. This shows that the combination of the CSA and Gibbs energy minimization engines such as {Thermo-Calc} has the potential to act as a reliable framework to accelerate the development of HEAs. However, this is dependent on the accuracy of the Thermo-Calc database. The TCHEA1 database was found to have a 70.8\% agreement with experimental data. Given the complexity of experimental studies of HEAs, it is difficult to determine whether the faults lie with Thermo-Calc in the form of thermodynamic or numerical errors, or with experimental methodologies such as not allowing enough time for thermodynamic equilibrium to occur. 

Future work may include collaborating with experimentalists to verify the single-phase solid solutions or precipitate hardenability regions predicted in this study, searching in new systems including those with more than five components, using site fraction ranges as constraints to determine degree of ordering, attempting to incorporate a kinetics model in the algorithm, and replacing the termination criterion with a method to maximize precision and recall.

\section{Acknowledgements}
T. K. acknowledges the support of the NSF through the project \emph{NRT-DESE: Data-Enabled Discovery and Design of Energy Materials (D$_3$EM)}, NSF-DGSE- 1545403. RA acknowledges the support of NSF through the project \emph{DMREF: Accelerating the Development of Phase-Transforming Heterogeneous Materials: Application to High Temperature Shape Memory Alloys}, NSF-CMMI-1534534. 

\section{Supplementary Material}

Additional figures (referred to in the manuscript) as well as the data set that lists the alloys as reported by Toda-Caraballo~\cite{toda2016criterion}, and their reported and predicted phase constitution is available at ~\cite{arroyave2018}.

\bibliographystyle{ActaMatnew-2}
\bibliography{ccsp_hea}

\end{document}